\documentclass[preprint]{aastex}

\slugcomment{To appear in The Astrophysical Journal Letters}

\shorttitle{Spin-induced precession in AGN}
\shortauthors{Caproni, Mosquera Cuesta \& Abraham}

\begin{document}


\title{Observational evidence of spin-induced precession in active galactic nuclei}


\author{Anderson Caproni\altaffilmark{1}, Herman J. Mosquera Cuesta 
\altaffilmark{2}, \and Zulema Abraham\altaffilmark{1} }
\altaffiltext{1}{Instituto de Astronomia, Geof\'\i sica e Ci\^encias
Atmosf\'ericas, Universidade de S\~ao Paulo, R. do
Mat\~ao 1226, Cidade Universit\'aria, CEP 05508-900, S\~ao Paulo, SP,
Brazil; acaproni@astro.iag.usp.br, zulema@astro.iag.usp.br.} 
\altaffiltext{2}{Instituto de Cosmologia, Relatividade e Astrof\'\i sica 
(ICRA-BR),  Centro Brasileiro de Pesquisas F\'\i sicas, R. Dr. Xavier 
Sigaud 150, CEP 22290-180, Rio de Janeiro, RJ, Brazil; hermanjc@cbpf.br.}








\begin{abstract}
We show that it is possible to explain the physical origin of  jet
precession in active galactic nuclei  (AGNs) through the misalignment
between the rotation axes of the accretion disk  and of the  Kerr black
hole.  We apply this scenario to  quasars, Seyfert galaxies and
also to the Galactic Center black hole Sgr\,A$^{\ast}$, for which signatures of 
either jet or disk
precession have been found. The formalism adopted is parameterized by
the ratio of the precession period to the black hole mass and can be
used to put constraints to the physical properties of the accretion disk
as well as to the black hole spin in those systems.
\end{abstract}




\section{Introduction}

   It is a well-accepted idea that the center of an AGN harbors a
supermassive black hole surrounded by  an accretion disk, the dynamics
of which eventually produces  a bipolar jet. In some AGNs, there were
found evidence of disk precession. In  Seyfert  galaxies this 
precession was detected, in direct form, from the variability of 
double-peaked Balmer lines   (e.g.,  \citealt{sto03}). In blazars, where
the  accretion disk emission is overrun  by the non-thermal jet
emission, precession can be inferred indirectly from the jet kinematics
(ejection of superluminal knots  in distinct directions and with
different apparent proper motions) or from periodic boosting of the 
continuum emission (e.g., \citealt{abra00,caab04a}). 

A question that could be arisen is: what is the  physical mechanism
driving the disk precession?  In Seyfert  galaxies, precession has been
attributed  to the relativistic advance of the pericenter in  an
elliptical accretion disk (e.g., \citealt{era95}),  to the precessing
single-spiral arm in a  circular disk (e.g., \citealt{sto03}), or to  a
radiation-induced warp in the disk (e.g.,  \citealt{prin96,sto97}).  For
blazars, several authors have proposed a binary black hole system
scenario  wherein the primary  accretion disk precesses due to torques
induced by the secondary black hole, whose orbital plane does  not
coincide with that of the accretion disk 
(\citealt{sil88,leva96,katz97,rom00,sti03,caab04a,caab04b}).  Although
supermassive binary black hole systems provide precession periods
compatible with  observations, the mean separation of the black holes
might be sometimes substantially short, so that  their timescale for
coalescence might reach less than  1000 years \citep{capr04}. 

In this work, we discuss the possibility that the disk precession arises
from torques induced by the  misalignment between the angular momenta of
a Kerr black hole and that of the accretion disk, which can  be applied
both to blazars and Seyfert galaxies. In $\S$ 2, we derive the relation
between precession  period, mass of the black hole and physical
parameters of the accretion disk. In $\S$ 3 we apply this  formalism to
several objects for which precession period and black hole mass are
known, and in $\S$ 4  we present our conclusions.  

\section{Spin-induced disk precession}

 Frame dragging produced by a Kerr black hole, known as Lense-Thirring
effect \citep{leth18}, leads a  particle to precess if its orbital plane
is inclined in relation to the equatorial plane of the black  hole (the
angular momentum of the black hole is perpendicular to this plane). The
amplitude of the precession angular velocity $\omega_{\mathrm{L-T}}$ 
decreases  with the cube of  the distance (e.g., \citealt{wilk72}) and
becomes negligible at large distances. 

If a viscous accretion disk is inclined in relation to the equatorial
plane of the Kerr black hole, the  differential precession will produce
warps in the disk.  The combined action of the Lense-Thirring effect and
the internal viscosity of the accretion disk  forces the alignment
between the angular momenta of the Kerr black hole and  the accretion
disk. This  is known as the Bardeen-Petterson effect \citep{bape75} and
affects only the innermost part of the  disk, while its outer parts tend
to remain in its original configuration, due to the short range of the 
Lense-Thirring effect.  The transition radius between these two regimes
is known as Bardeen-Petterson radius, and its exact location depends on
the Mach number of the accretion disk
(\citealt{bape75,kupr85,scfe96,ivil97,nepa00}).
 
   \citet{nepa00} studied numerically the time evolution of an accretion 
   disk due to the Bardeen-Petterson effect produced by a maximally rotating 
   black hole, and found that the initial differentially precessing disk 
   evolves towards a rigid-body-like precession. Thus, we will assume 
   hereafter that the accretion disk precesses as a rigid-body.

   In the reference frame precessing with the accretion disk, the total
torque $\vec{T}$ due to the  Bardeen-Petterson effect acting     upon
the misaligned accretion disk is given by \citep{pate95}:

   \begin{equation}
 \vec{T} = 
\frac{d\vec{L}_\mathrm{d}}{dt}+\omega_\mathrm{prec}\frac{\vec{J}_\mathrm{BH}}
 {\vert\vec{J}_\mathrm{BH}\vert}\times\vec{L}_\mathrm{d}, 
   \end{equation}    
   \\where $\vec{L}_\mathrm{d}$ and $\omega_\mathrm{prec}$ are
respectively the total angular momentum  of the disk and the precession
angular velocity, $\vec{J}_\mathrm{BH}$ is the angular momentum of the 
black hole, defined as $\vert\vec{J}_\mathrm{BH}\vert =
GM_\mathrm{BH}^2\vert a_\ast\vert/c$,  $G$ is  the gravitational
constant, $M_\mathrm{BH}$ the mass of the black hole and $c$  the light
speed.  The dimensionless parameter $a_\ast$ ($-1 \leq a_\ast \leq 1$)
is the ratio  of the angular momentum of  the compact object and that of
a Kerr black hole rotating at its maximal velocity.  The signs "+" and 
"-" correspond respectively to prograde and retrograde black hole
rotation relative  to the angular  velocity of the accretion disk.

 In the Kerr metric, the radius of the marginally stable orbit
$R_\mathrm{ms}$, which  will be taken as  the inner radius of the
accretion disk,  is given by $R_\mathrm{ms} = \xi_\mathrm{ms}R_\mathrm{g}$, 
where $R_\mathrm{g} = GM_\mathrm{BH}/c^2$ is the gravitational radius, 
$\xi_\mathrm{ms} = 3+A_2\mp\sqrt{(3-A_1)(3+A_1+2A_2)}$, with 
$A_1 = 1+(1-a_\ast^2)^{1/3}\left[(1+a_\ast)^{1/3}+(1-a_\ast)^{1/3}\right]$ 
and $A_2 = \sqrt{3a_\ast^2+A_1^2}$. In this case, the minus and plus
signs correspond to prograde and  retrograde motion, respectively. 
 
When the system reaches a quasi-stationary state, the precession period
of the disk,   $P_\mathrm{prec}$, is given in the source's reference
frame by \citep{lime02}: 

   \begin{eqnarray}
      P_\mathrm{prec} = 2\pi\sin\theta\left(\frac{L_\mathrm{d}}{T}\right),
   \end{eqnarray}   
   \\where $\theta$ is the angle between the orientations of the angular
momenta of the black hole and  that of the accretion disk. Assuming that
the surface density of the disk $\Sigma_\mathrm{d}$ depends  only on the
radius $r$  and that there is no warping in the disk, $L_\mathrm{d}$ can
be written as: 

   \begin{eqnarray}
      L_\mathrm{d} = 2\pi\int_{R_\mathrm{ms}}^{R_\mathrm{out}}
      \Sigma_\mathrm{d}(r)
      \Omega_\mathrm{K}(r)r^3 dr,
   \end{eqnarray}
   \\where $R_\mathrm{out}$ is the outer radius of the precessing part of 
   the disk and $\Omega_\mathrm{K}$ is the relativistic Keplerian 
   angular velocity, given as:

   \begin{eqnarray}
\Omega_\mathrm{K}(r) =
 \frac{c^3}{GM_\mathrm{BH}}\left[\left(\frac{r}{R_\mathrm{g}}
 \right)^{3/2}+a_\ast \right]^{-1}.
   \end{eqnarray}

   For small precession angles $\theta$,  $T$ can be represented by 
   \citep{lime02}:

   \begin{eqnarray}
      T=
	4\pi^2\sin\theta\int_{R_\mathrm{ms}}^{R_\mathrm{out}}
	\Sigma_\mathrm{d}(r)\Omega_\mathrm{K}(r)
	\nu_\mathrm{p\theta}(r)r^3dr,
   \end{eqnarray}   
   \\with the nodal precession frequency $\nu_\mathrm{p\theta}$ given by 
   \citep{kato90}:

   \begin{eqnarray}
      \nu_\mathrm{p\theta} =
 \frac{\Omega_\mathrm{K}(r)}{2\pi} \left[1 - \sqrt{1-4a_\ast 
\left(\frac{R_\mathrm{g}}{r}
 \right)^{3/2}
 + 3a_\ast^2 \left(\frac{R_\mathrm{g}}{r}\right)^{2}}\right].
   \end{eqnarray}

   Substituting equations [4] and [6] into equations [3] and [5], and introducing the 
   dimensionless variable $\xi=r/R_\mathrm{g}$, we obtain:

   \begin{eqnarray}
      P_\mathrm{prec} = \biggl(\frac{2\pi
	   GM_\mathrm{BH}}{c^3}\biggr)\frac{\int_{\xi_\mathrm{ms}}^{\xi_\mathrm{out}}
\Sigma_\mathrm{d}(\xi)\left[\Upsilon(\xi)\right]^{-1}\xi^3d\xi}
{\int_{\xi_\mathrm{ms}}^
{\xi_\mathrm{out}}\Sigma_\mathrm{d}(\xi)\Psi(\xi)\left[\Upsilon(\xi)\right]^{-2}\xi^3
d\xi},
   \end{eqnarray}
      \\with $\xi_\mathrm{out} = R_\mathrm{out}/R_\mathrm{g}$, $\Upsilon(\xi) =
    \xi^{3/2}+a_\ast$ and  $\Psi(\xi) = 
1-\sqrt{1-4a_\ast\xi^{-3/2}+3a_\ast^2\xi^{-2}}$.
   
   This equation is parametrized by the relation between the precession period and 
the 
   black hole mass, which depends  on the black hole spin and on how the disk surface 
density  varies with  radius. In this work, we 
   shall adopt two different descriptions for this last parameter: a power-law 
function, such 
   as $\Sigma_\mathrm{d}(\xi)=\Sigma_0\xi^s$ (e.g., 
   \citealt{pate95,larw97,nepa00}), and an exponential function 
   $\Sigma_\mathrm{d}(\xi)=\Sigma_0 e^{\sigma\xi}$ (e.g., 
   \citealt{stsp01,dan03}), where $\Sigma_0$, $s$ and $\sigma$ are 
   arbitrary constants.

\section{Predictions from the spin-induced precession model for an AGN sample}

   In Figure 1, we show the iso-contours of the ratio $P_\mathrm{prec}/M_\mathrm{BH}$ 
   calculated from equation [7] in the [log$ (R_\mathrm{out}/R_\mathrm{ms})$, 
   $a_\ast$]-plane. The integrals in that equation  were calculated for values 
of $R_\mathrm{out}$ between 1.1 and  1000$R_\mathrm{ms}$.
\footnote{The extent of the outer edge of an accretion disk is controversial. 
   For instance, \citet{codu90} assumed an accretion disk that extends up to about 
   10$^5R_\mathrm{g}$, while \citet{good03} argue that, due to self-gravity, the 
   outer parts of the disk might fragmentise, reducing its size to about  
2000$R_\mathrm{g}$.} 
   
   The calculations were performed for values of $a_\ast$ between -1 and 1.
   \footnote{Although \citet{thor74} has shown that $a_\ast\leq 0.998$ 
   for a thin Keplerian disk due to the photon capture by the black hole, 
   the black hole can spin up either by black hole mergers \citep{agkr00}, 
   or through accretion from a thick disk \citep{abla80}. } We have considered 
   both power-law and exponential profiles for the surface density using 
   four different values of $s$ (-2, -1, 0 and 2) and for $\sigma$ 
   (-2, -1, 0 and 0.05). \footnote{Accretion disk models found in the 
   literature, such as those developed by \citet{shsu73}, usually provide 
   a radial dependency of the surface density in this range.} 

\begin{figure*}
      {\centerline{\epsscale{0.9}\plotone{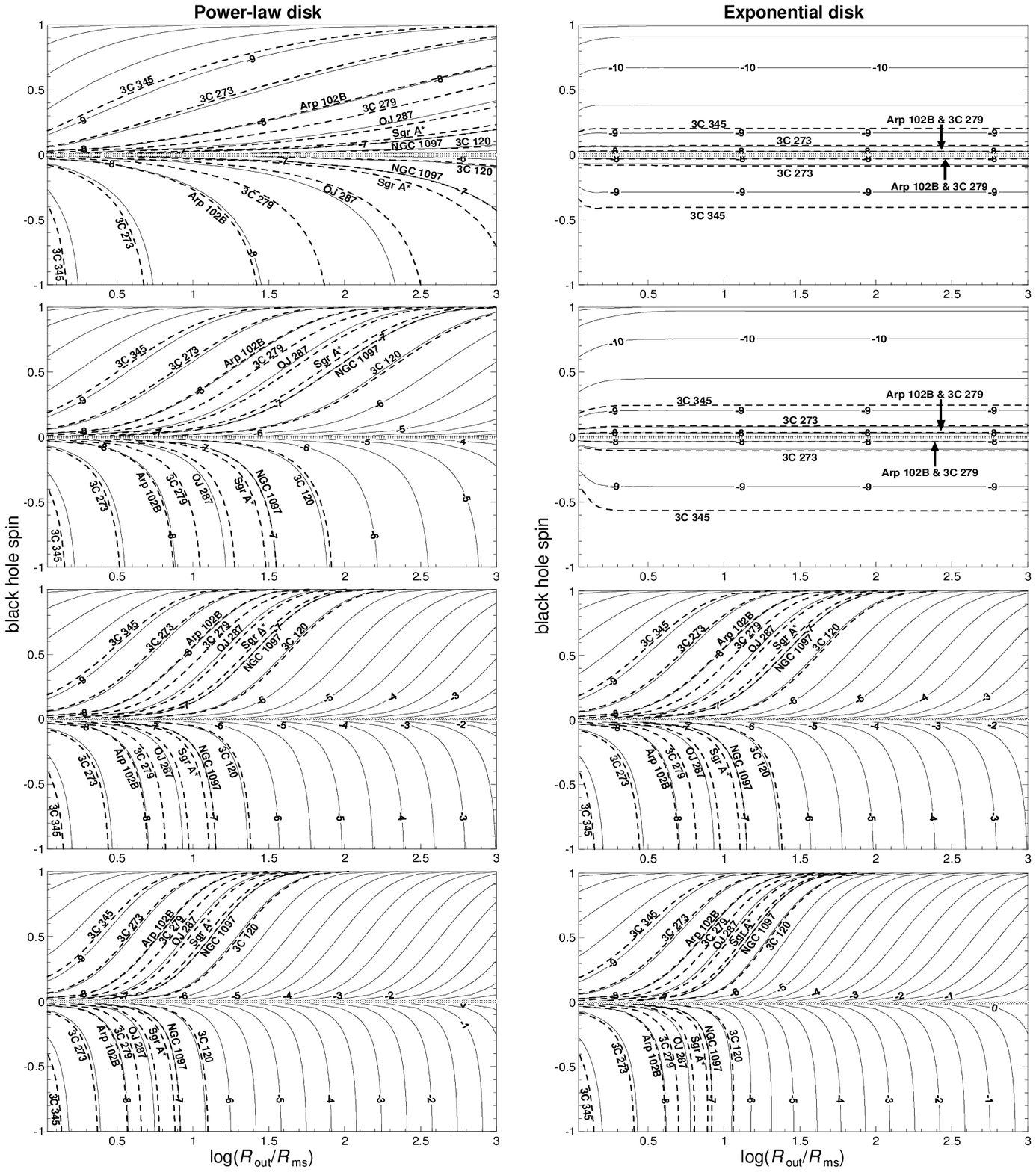}}}
\caption{Iso-contours (in logarithm scale) of  the ratio between the disk 
precession period and the black hole mass
(in units of yr M$_\sun^{-1}$) in the plane defined by the black hole spin  and the 
disk outer radius,  for a spin-induced precession  model. Left panel
corresponds to a power law surface  density profile ($s$=-2,                
-1, 0 and 2, from top to bottom) and  right panel to an                 
exponential disk ($\sigma$=-2, -1, 0 and 0.05, from top to bottom).                
The contours of the eight sources studied in this work are represented                
by dashed lines.  The parameter range $-0.01\leq a_\ast \leq 0.01$, for which  
 the solutions for the outer radius become very small, were  excluded from our 
calculations.}
         \label{Precession period}
   \end{figure*}

   \placefigure{Precession period}

   One can see that the general behaviour of 
   $P_\mathrm{prec}/M_\mathrm{BH}$ depends strongly on the  accretion
   disk structure ($s$ or $\sigma$), specially in the case of
   exponential  disks. Indeed, if we maintain $a_\ast$ fixed, the
   increase of  $P_\mathrm{prec}/M_\mathrm{BH}$ with $R_\mathrm{out}$
   occurs at a higher rate as  $s$ or $\sigma$ increases. In some cases
   when $\sigma<0$,  $P_\mathrm{prec}/M_\mathrm{BH}$ tends
   asymptotically to a  constant value that, on the other hand, will be
   reached at lower values of $R_\mathrm{out}$ and $\vert a_\ast \vert$ 
   as $\sigma$ becomes more
   negative.  This is a consequence of the rapid decrease of the  amount
   of disk material with the distance in such  configurations. Indeed,
   this trend is also seen in  power-law disks with $s<-2$.

   For a particular $R_\mathrm{out}/R_\mathrm{ms}$, the same value of   
   $P_\mathrm{prec}/M_\mathrm{BH}$ is obtained for a smaller absolute    
   value of $a_\ast$ in the case of prograde rotation, giving an    
   asymmetry in the curves shown in Figure 1. This can be explained if we
   consider that in this case,  the inner disk boundary is closer to the
   gravitational radius, where the Lense-Thirring effect is more 
   prominent. 
   
   To compare the model predictions with observations we  analysed a 
   set of sources for which information of both precession periods and 
   black hole masses are available. The selected sources  correspond 
   to the Seyfert I galaxies NGC\,1097 and 3C\,120, the broad-line 
   radio galaxy Arp\,102B, the BL Lac OJ\,287, the quasars 3C\,273, 
   3C\,279 and 3C\,345, and the Galactic Center source Sgr\,A$^\ast$ 
   \footnote{Although our  Galaxy is not exactly an AGN, the supermassive black 
   hole in its center is a good 
   candidate for this study since there are already a large number of 
   observational and theoretical investigations on this source (e.g., 
   \citealt{mel01}, \citealt{lime02}, \citealt{lime03}). Hence the 
   validity of the model may be tested by combining the results here 
   with other aspects of the source.}, whose parameters and references 
   are listed in Table 1.

   \placetable{tbl-1}
 
   The dashed lines in Fig. 1 correspond to the values of 
   $P_\mathrm{prec}/M_\mathrm{BH}$ for each object listed in Table 1.
   One can  note that there is always a set of parameters 
   ($a_\ast$, $s/\sigma$, $R_\mathrm{out}$) capable of reproducing 
   the precession period related to the sources in both disk models, 
   indicating that the spin-induced precession can be a feasible 
   engine for disk/jet precession. However, not all the disk models 
   and sizes can provide precession periods as those found in our 
   sample of AGN. This can put constrains on the physical parameters of 
   the disks present in the AGNs for which evidence of precession 
   were found. For instance, $P_\mathrm{prec}/M_\mathrm{BH}$ of the 
   quasar 3C\,345 is reproduced by the spin-induced precession model 
   in the case of a power-law disk with $s=-2$ only if the black hole 
   has a spin  $a_\ast\leq -0.35$ or  $a_\ast\geq 0.2$.

\section{Conclusions}

    In this work we analyzed the possibility of jet/disk precession  in
AGNs    be driven by torques in an accretion disk produced by the
misalignment     between the angular momenta of the black hole and that
of the accretion disk.     We adopted a formalism based on
\citet{lime02}, and obtained an equation  for the ratio of the 
precession period to the black hole mass, parametrized by the Kerr
metric and the disk structure, which  is mostly determined by  physical
processes, such as viscosity and radiation. This expression can be  used 
to extract the physical properties of the     accretion disk, as well as
of the black hole spin in astrophysical     systems that present
signatures of disk/jet precession.

   Two different disk models were used in the calculations: power law and  
exponential surface density profiles. We applied
these models to eight  objects for which precession periods and black
hole  masses are published. We  showed that it is always possible to
obtain a set of parameters that provides a jet/disk precession
period compatible with those derived from high-resolution radio maps    
and continuum variability (e.g., \citealt{lime02,caab04b}), or from the    
double-peak emission-line variability (e.g., \citealt{new97,sto03}).    
However, not all the possible combinations of $a_\ast$, $s/\sigma$ and    
$R_\mathrm{out}$ render results compatible with the observation, putting
 constraints  on the disks  properties if the precession is due 
to the proposed mechanism. 

\acknowledgments

   This work was supported by the Brazilian Agencies CNPq, FAPESP and FAPERJ.
   We would like to thank the anonymous referee for the useful commentaries
   and suggestions.

\clearpage


\onecolumn

\clearpage


\begin{deluxetable}{ccccccc}
\tabletypesize{\scriptsize}
\tablecaption{Physical parameters of the sources. \label{tbl-1}}
\tablewidth{0pt}
\tablehead{
\colhead{Object} & \colhead{z} & \colhead{$P_\mathrm{prec}^\mathrm{obs}$ 
(yr)}\tablenotemark{a} & \colhead{Ref.} & \colhead{$M_\mathrm{BH}$ (M$_\sun$)} 
& \colhead{Ref.} & \colhead{$P_\mathrm{prec}/M_\mathrm{BH}$ (yr M$_\sun^{-1}$)} 
}
\startdata
Sgr\,A$^\ast$ & 0.000 & 0.29 & (1) & 3.70$\times$10$^{6}$ & (9)  & 
7.84$\times$10$^{-8}$\\
NGC\,1097     & 0.004 & 5.5  & (2) & 5.50$\times$10$^{7}$ & (2)  & 
9.96$\times$10$^{-8}$\\
Arp\,102B     & 0.024 & 2.2  & (3) & 2.20$\times$10$^{8}$ & (3)  & 
9.59$\times$10$^{-9}$\\
3C\,120       & 0.033 & 12.3 & (4) & 3.40$\times$10$^{7}$ & (10) & 
3.50$\times$10$^{-7}$\\
3C\,273       & 0.158 & 16.0 & (5) & 4.93$\times$10$^{9}$ & (11) & 
2.80$\times$10$^{-9}$\\
OJ\,287       & 0.306 & 11.6 & (6) & 2.29$\times$10$^{8}$ & (12) & 
3.88$\times$10$^{-8}$\\
3C\,279       & 0.536 & 22.0 & (7) & 8.17$\times$10$^{8}$ & (11) & 
1.75$\times$10$^{-8}$\\
3C\,345       & 0.593 & 10.1 & (8) & 7.96$\times$10$^{9}$ & (11) & 
7.97$\times$10$^{-10}$\\
\enddata
\tablecomments{$z$ is the redshift and $P_\mathrm{prec}^\mathrm{obs}$ 
is the precession period measured in the observer's reference frame 
[$P_\mathrm{prec}^\mathrm{obs}=(1+z)P_\mathrm{prec}$].}
\tablerefs{(1) \citealt{zha01}; (2) \citealt{sto03}; (3) \citealt{new97}; (4) 
\citealt{caab04b}; (5) \citealt{abro99}; (6) \citealt{abra00}; (7) \citealt{abca98};
(8) \citealt{caab04a}; (9) \citealt{sch02}; (10) \citealt{pet98}; (11) 
\citealt{gu01};
(12) \citet{xie02}.}

\end{deluxetable}
%
%

\end{document}